\documentclass[%
reprint,
amsmath,amssymb,
noeprint,
aip,
]{revtex4-2}

\usepackage{graphicx}
\usepackage{dcolumn}
\usepackage{bm}
\usepackage{amsmath}
\usepackage{hyperref}

\begin{document}
	
	\preprint{APS/123-QED}
	
	\title{Observations about utilitarian coherence in the avian compass}
	
	\author{Luke D. Smith}
	\author{Jean Deviers}
	\author{Daniel R. Kattnig*}%
	\affiliation{%
		Living Systems Institute and Department of Physics \\ University of Exeter, Stocker Road, Exeter EX4 4QD, United Kingdom \\
		* Corresponding author: d.r.kattnig@exeter.ac.uk
	}%
	
	\date{\today}
	
	\begin{abstract}
    	\textbf{\abstractname.}  It is hypothesised that the avian compass relies on spin dynamics in a recombining radical pair. Quantum coherence has been suggested as a resource to this process that nature may utilise to achieve increased compass sensitivity. To date, the true functional role of coherence in these natural systems has remained speculative, lacking insights from sufficiently complex models. Here, we investigate realistically large radical pair models with up to 21 nuclear spins, inspired by the putative magnetosensory protein cryptochrome. By varying relative radical orientations, we reveal correlations of several coherence measures with compass fidelity. Whilst electronic coherence is found to be an ineffective predictor of compass sensitivity, a robust correlation of compass sensitivity and a global coherence measure is established. The results demonstrate the importance of realistic models, and appropriate choice of coherence measure, in elucidating the quantum nature of the avian compass.
	\end{abstract}
	
	\maketitle
	
	
\section*{Introduction}
\noindent Quantum coherence and entanglement are central enablers of quantum technologies and quantum information processing. The emerging field of quantum biology \cite{Ball2011,Lambert2013,McFadden2018,Marais2018,Kim2021} hypothesizes that these rudiments of quantumness could likewise empower biological processes of living systems. In particular, long-lived quantum coherence is an overarching theme of many of the focus points of quantum biology, such as the avian magnetoreceptor, and photosynthetic energy transport \cite{Romero2014,Kominis2015,Duan2017,Thyrhaug2018,Cao2020}. However, our understanding and quantification of coherence as an operational resource \cite{Baumgratz2014,Streltsov2016,Winter2016}, and its role in the complex quantum systems of life, is still limited.
	
\begin{figure*}[t]
	\centering
		\includegraphics[width=0.8\linewidth]{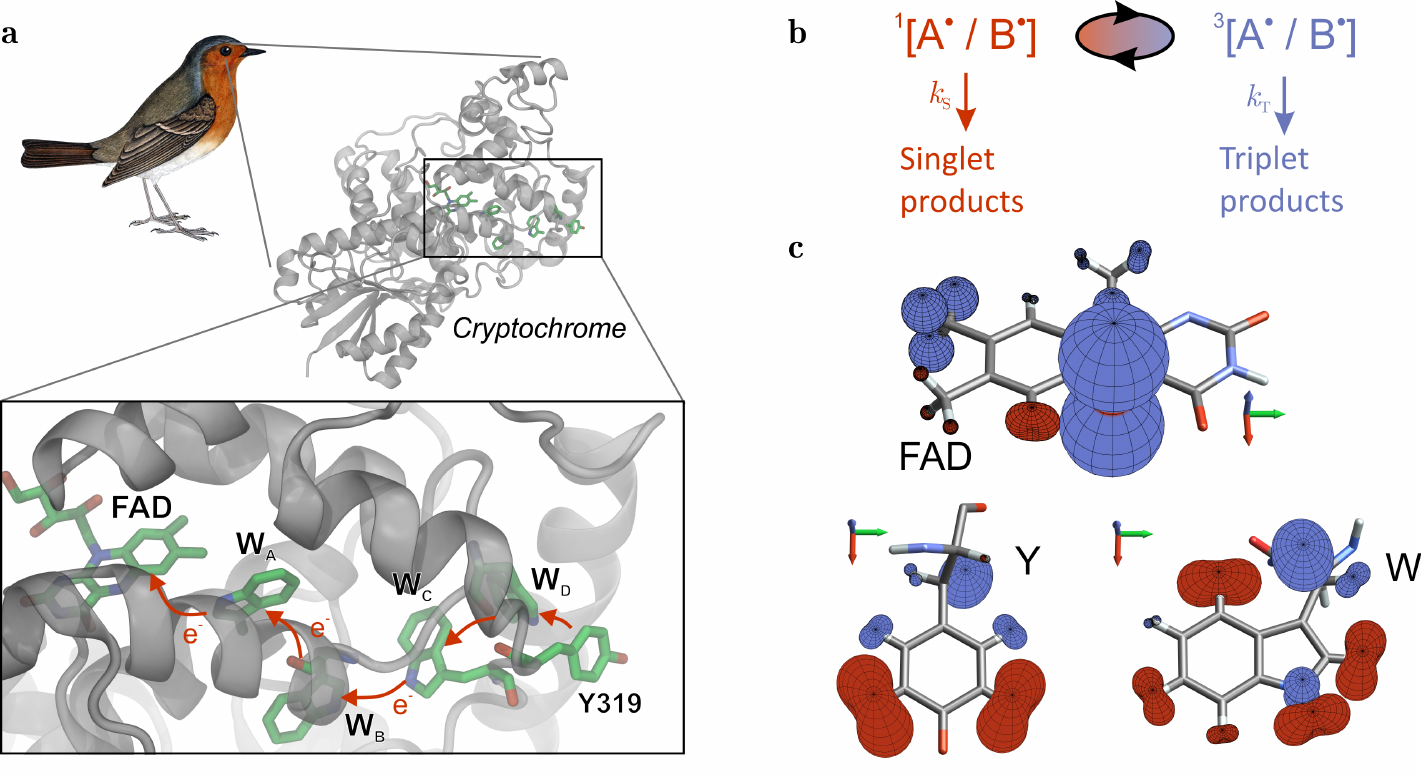}
		\caption{\label{fig:scheme}\textbf{a} Structure of an avian cryptochrome (PDB identifier: 6PU0, \textit{Columba livia}), including the central electron transfer chain comprising four tryptophans (W) labelled A (W395), B (W372), C (W318) and D (W369) and ending in the surface exposed tyrosine Y319. Photo-excitation of FAD in cryptochrome initiates consecutive electron transfer reactions of adjacent donors/acceptor pairs (red arrows), producing sequential radical pairs of the form [FAD$^{\bullet-}$ / W$^{\bullet+}$] and possibly [FAD$^{\bullet-}$ / Y$^{\bullet}$] (see e.g.\ ref.\ \onlinecite{Xu2021}). The well separated radical pairs involving W$_{C}$ and W$_{D}$ have been implicated with magnetoreception. Alternative radical pair models have been discussed, e.g., in the context of dark-state reoxidation \cite{P2011, Wiltschko2016, Player2019}. \textbf{b} Generic radical-pair reaction scheme. The radical pair is born in the singlet state (spin multiplicity indicated by superscript labels, total electronic spin: $S = 0$) which, via coherent interconversion, can interconvert to a triplet state ($S = 1$). Here, A$^{\bullet}$ is assumed to be FAD and B$^{\bullet}$ is a tryptophan (W) or tyrosine (Y) residue. The radical pair may form singlet and triplet products with rate constants $k_{S}$ and $k_{T}$, respectively. The singlet channel typically comprises radical pair recombination and spin-independent protein structural rearrangements; the latter also contributes to the triplet channel. \textbf{c} Graphical representation of hyperfine interactions in FAD$^{\bullet-}$ (top), Y$^{\bullet}$ (bottom left) and W$^{\bullet+}$ (bottom right). Here, in the direction given by the unit vector $\mathbf{d}$, the plotted surfaces are drawn at distance $\vert \vert \mathbf{Ad} \vert \vert$ from the location of the nucleus in question, whereby $1\mskip3mu$\r{A} corresponds to $17\mskip3mu$MHz. Surfaces are coloured according to the sign of the projection, with blue and red corresponding to positive and negative signs, respectively. The molecules are shown in their respective standard orientations.}
	\end{figure*}

	The avian compass and several related magnetosensitive feats are hypothesised to rely on coherent spin dynamics in a radical pair \cite{Hore2016} (RP), putatively formed in the blue-light sensitive flavo-protein cryptochrome. The radical pair comprises two unpaired electrons, the combined spin angular momentum of which are described in terms of singlet and triplet states. Typically, only the singlet state can recombine to re-form the diamagnetic resting state, while both singlets and triplets can give rise to spin-independent structural rearrangements of the protein and thus signal to down-stream processes (see the scheme presented in Fig. \ref{fig:scheme}) \cite{Ritz2000}. Magnetosensitivity emerges in these settings from a change in singlet and triplet state populations and thus yields for different magnetic field orientations. This is driven by the coherent singlet-triplet interconversions which result from the hyperfine coupling with magnetic nuclei in the radicals under perturbation by the comparably weak Zeeman interaction of the electron magnetic moments with the geomagnetic field ($\approx 50\mskip3mu\mathrm{\mu}$T). The nonstationary singlet-triplet coherence is an essential requirement for this process. In its absence (i.e., if singlet and triplet states are eigenstates of the Hamiltonian), no interconversion between singlet and triplet states occurs and the reaction yield will be insensitive to the applied magnetic field. On the other hand, electronic spin state decoherence, through interaction with the environment of nuclear spins, provides a natural companion to the radical pair spin dynamics and prerequisite for its sensitivity to magnetic fields \cite{Tiersch2012}. In order to elicit sensitivity to a $50\mskip3mu\mathrm{\mu}$T field the spin coherence must persist for about a microsecond, i.e., the reciprocal of the electron Larmor precession frequency ($\approx 1.4\mskip3mu$MHz), or longer. Longer-lived singlet-triplet coherences promise a more precise magnetic field measurement, which could underpin a compass of exquisite acuity \cite{Hiscock2016}. However, a coherent lifetime of the radical pair of $1$--$10\mskip3mu\mathrm{\mu}$s has been deemed realistic based on predictions of spin relaxation processes and animal behavioural experiments \cite{Kobylkov2019,Kattnig2016b,Kattnig2016a}.
	Current models of magnetoreception implicate the protein cryptochrome as the host of the magnetosensitive radical pair \cite{Ritz2000}. A photo-generated radical pair comprising a flavin anion radical (FAD$^{\bullet-})$ and radical cation derived from a surface-exposed tryptophan residue (W$^{\bullet +}$) is a popular model supported by \textit{in vitro} studies \cite{Xu2021}. Alternative radical pair models have been put forward based on \textit{in vivo} observations and modelling \cite{Atkins2019, P2011,Lee2014,Wiltschko2016,Player2019,Procopio2020} (e.g. the flavin semiquinone/superoxide pair). Here, we focus on the former and an additional model comprising a tyrosine radical partner instead of the tryptophan \cite{Atkins2019}.
	
	Several works have suggested that coherence is a resource for the sensitivity of the compass, insofar as a larger coherence corresponds to increased directional magnetic sensitivity. Cai and Plenio \cite{Cai2013} have analysed many randomly chosen prototype radical pair systems (with 5--6 nuclear spins) and found that their ``global coherence'' of the electron and nuclear spin system is a predictor of compass sensitivity. Kominis has introduced a formal measure of singlet-triplet coherence \cite{Kominis2020} and demonstrated that singlet-triplet coherence of radical pairs provides an operational advantage to magnetoreception in simple model systems involving one nuclear spin. In a previous study \cite{Atkins2019}, we have observed that the electron coherences in realistic multi-nuclear radical pairs comprising up to 21 nuclear spins are long-lived compared to the electronic entanglement. We concluded that the compass is coherence-driven while entanglement does not confer an operational advantage of the compass. The latter point has also been elaborated on in ref.\ \onlinecite{Hogben2012} with comparable conclusions. Le and Olaya-Castro have recently investigated a three-spin system subject to collisional, symmetry-breaking environmental interactions \cite{Le2020}. The authors demonstrate that magnetosensitivity in this system requires coherence and suggest that a small degree of coherence, regardless of basis, is likely a quantum resource for biomolecular systems.
	
	The property of coherence as a facilitator of increased compass sensitivity was recently questioned by Jain \textit{et al}., who suggest that the electronic coherence swiftly decays with the number of coupled nuclei. The authors demonstrate that the compass can provide high sensitivity despite operating in a parameter regime without ``sustained'' electron spin coherence \cite{Jain2021}. This conclusion was derived by comparing systems with up to 6 magnetic nuclei coupled with identical, aligned and axial hyperfine tensors (i.e., with identical transverse and dominant longitudinal components). While this system shows sustained electron coherences for vanishing transverse hyperfine components, the system with maximal sensitivity did not sustain the coherences (with “sustained” construed in relation to the artificial reference system). As of now, it is not clear if the findings of previous coherence measure studies translate to more realistic conditions or models, but the conclusion of Jain \textit{et al.} is certainly surprising, as it is in stark deviance to the abovementioned studies.
	
	Here, we investigate the utilitarian character of coherence for the compass based on models that are biologically relevant. In particular, we focus on radical pairs that have been implicated with cryptochrome-magnetoreception, namely flavin/tryptophan and flavin/tyrosine radical pairs with up to 21 nuclear spins. Our study thus endeavours to address the character of coherence in the actual magnetoreceptor, as currently envisaged, rather than the principle role as elaborated in previous studies using strongly simplified/abstract models. Instead of varying the hyperfine interaction parameters, which are predetermined by the identity of the radical pair and only weakly dependent on the molecular environment \cite{Procopio2016}, we vary the relative orientation of the two radicals. This is a parameter which could have been subject to evolutionary optimisation in the protein cryptochrome and is more likely to reflect any potential quantum advantage that is exercisable in the well-defined biological system, if it exists. Specifically, considering more realistic modelling conditions, we analyse whether the relative orientations of optimal compass fidelity are correlated with the electronic coherence (as anticipated based on investigations \cite{Cai2013,Kominis2020,Atkins2019}) or not (as advocated in a recent study by Jain \textit{et al.} \cite{Jain2021}). Our results question the effectiveness of coherence, measured by electronic coherence quantifiers, as the sole driver of magnetoreception and explanation of increased compass sensitivity. However, global coherence is realized as a persistent effectuator of compass fidelity, confirming the hypothesized quantum nature of the processes even for realistic systems with relatively short radical lifetimes of $1\mskip3mu\mathrm{\mu}$s.

	\section*{\label{sec:level1}Results}
	
	\noindent \textbf{Radical pair model}. We consider radical pairs subject to coherent evolution under the local Hamiltonians of their constituent radicals and spin-selective recombination proceeding with the same rate constant, $k_{S}=k_{T}=k$, for the singlet and triplet channel. Inter-radical interactions are neglected, as is spin relaxation. In this simplified, but commonly studied scenario \cite{Hiscock2016,Atkins2019,Lee2014}, the spin dynamics of the RP and its associated magnetic field effects (MFEs) can be described by the following master equation for the spin density operator $\hat{\rho}(t)$:
	\begin{align}
			\frac{{{\rm{d}}\hat \rho (t)}}{{{\rm{d}}t}} &=  - i\left[ {\hat H,\hat \rho (t)} \right] - \frac{{{k_S}}}{2}\left\{ {{{\hat P}_S},\hat \rho (t)} \right\} - \frac{{{k_T}}}{2}\left\{ {{{\hat P}_T},\hat \rho (t)} \right\} \nonumber \\
			&=  - i\left[ {\hat H,\hat \rho (t)} \right] - k\hat \rho (t), \label{eq.master_equation}
	\end{align}
	where [] and \{\} denote the commutator and anti-commutator, respectively. Here, the Hamiltonian is of the form $\hat H = {\hat H_{\rm{A}}} + {\hat H_{\rm{B}}}$ with subscripts A and B referring to the two radicals, and $\hat{P}_{S}$ and $\hat{P}_{T} = \hat{1} - \hat{P}_{S}$ are projection operators onto the singlet state and triplet manifold, respectively. The chemical reactivity has been included using the Haberkorn approach \cite{Haberkorn1976}, which gives rise to minimal singlet-triplet dephasing (alternative approaches have been suggested, but for $k_{S}=k_{T}$ the differences of various master equation are lessened \cite{Kominis2020,Fay2018}). The Hamiltonians $\hat{H}_{i}$, $i \in \{A, B\}$, account for the Zeeman interactions between the electron spins and the applied magnetic field and the hyperfine interactions between the electron and nuclear spins within radical $i$ ($\hbar  = 1$):
	\begin{align}
		{\hat H_{\rm{i}}} = \sum\limits_j {{{{\bf{\hat S}}}_i} \cdot {{\bf{A}}_{i,j}} \cdot {{{\bf{\hat I}}}_{i,j}}}  + {\vec \omega _i} \cdot {{\bf{\hat S}}_i}.
	\end{align}
	Here, the Larmor precession angular frequency is given by $\vec{\omega}_{i} = -\gamma_{i}\vec{B}$, with $\gamma_{i}$ denoting the gyromagnetic ratio of the electron in radical $i$ and $\vec{B}$ the applied magnetic field. $\mathbf{A}_{i, j}$ is the hyperfine coupling tensor between the $j$th nuclear spin and the $i$th electron spin; $\hat{\mathbf{I}}_{i,j}$ and $\hat{\mathbf{S}}_{i}$ are the corresponding vector operators of nuclear and electron spin angular momentum.
	
	Here, we solve Eq.\ (\ref{eq.master_equation}) for realistically complex spin systems based on an approach employing spin correlation tensors (see Methods for details) \cite{Atkins2019,Manolopoulos2013,Schulten1978}, to yield the quantum yield of recombination via the singlet channel, $Y_{S}(\vartheta,\varphi)$. Its sensitivity to changes in the applied magnetic field direction determine the compass fidelity, which we here assess in terms of established measures, namely the absolute anisotropy defined as
	\begin{align}
		\Delta_{S} = \max_{\vartheta, \varphi}Y_{S}(\vartheta,\varphi)-\min_{\vartheta, \varphi}Y_{S}(\vartheta,\varphi),
	\end{align}
	where the polar angle $\vartheta$ and azimuthal angle $\varphi$ parametrize the direction of the applied magnetic field in the protein frame. We will relate this sensitivity measure to quantifiers of the quantum coherence present in the radical pair during its magnetosensitive spin evolution.
	
	\
	
	\noindent \textbf{Measures of coherence}. Various measures to quantify coherence have been developed \cite{Baumgratz2014,Winter2016,Streltsov2016}. Several such measures have previously been used in the context of the avian compass \cite{Cai2013,Kominis2020,Gauger2011,Pauls2011,Jain2021}. One canonical measure to quantify coherence is the relative entropy of coherence, which, in terms of the normalized density operator $\hat{\rho}$ and a chosen basis, $\{\vert n \rangle \}_{n=1}^{d}$, of the $d$-dimensional Hilbert space, is defined as
	\begin{align}
		\mathcal{C}_{r}[\hat{\rho}]=S[\mathbb{I}\mathbb{C}(\hat{\rho})] - S[\hat{\rho}], \label{eq.rel_entropy}
	\end{align}
	where $S[\hat{\rho}] = - \mathrm{Tr}[\hat{\rho}\log(\hat{\rho})]$ is the von Neumann entropy and $\mathbb{IC}$ denotes the dephasing operation $\mathbb{IC}(\hat{\rho}) = \sum_{n} \vert n \rangle \langle n \vert \hat{\rho} \vert n \rangle \langle n \vert$, which maps any quantum state into an incoherent state in the chosen basis.
	
	Another measure, the $l_{1}$-norm of coherence \cite{Baumgratz2014} is defined as 
	\begin{align}
		\mathcal{C}_{l_{1}}[\hat{\rho}] = \sum_{n\neq m} \vert \langle n \vert \hat{\rho} \vert m \rangle \vert. \label{eq.l1_norm}
	\end{align}
	Furthermore, we employ an additional measure suggested by Kominis defined by
\begin{align}
	\mathcal{C}_{st}[\hat{\rho}] = S[\hat{P}_{S}\hat{\rho}\hat{P}_{S}+\hat{P}_{T}\hat{\rho}\hat{P}_{T}] - S[\hat{\rho}],
\end{align}
which reports singlet-triplet coherence while being independent of basis and unaffected by the degree of triplet coherence, which the author suggests was superior in assessing coherence in the radical pair-based compass \cite{Kominis2020}. Similarly, an alternative $l_{1}$-based measure of singlet-triplet coherence was used in ref.\ \onlinecite{Kritsotakis2014}. The relative entropy of $\hat{\rho}$ with respect to the maximally mixed state, $S\left[ {\hat \rho ||\hat 1/d} \right] = \log d - S\left[ {\hat \rho } \right]$, has likewise been employed to quantify the amount of basis-independent coherence \cite{Le2020}.
	Several alternative measures have been discussed in the literature \cite{Kritsotakis2014, Kominis2020, Le2020}. We provide a comprehensive collection of results for these alternatives in the Supporting Information (SI), but focus on the above-mentioned measures in the main text for conciseness.

	The coherence quantifiers introduced above can be employed to the entire density operator, $\hat \rho $, as well as to the electronic part, $\hat \sigma $, and are either basis-independent or defined with respect to a basis. In the present study, two clear choices include the singlet-triplet basis and the up-down basis of the electronic subspace, both of which will be used to evaluate $\mathcal{C}_{r}$ and $\mathcal{C}_{l_{1}}$ below. To distinguish these choices, we will henceforth index the coherence labels with $\mathcal{G}$ (global, based on $\hat{\rho}$), $\mathcal{E}$ (electronic, based on $\hat{\sigma}$), and ST (singlet-triplet basis) and UD (up-down basis). For example, $\mathcal{C}_{r}^{\mathcal{E},U\!D}$ corresponds to the electronic coherence assessed via the relative entropy quantifier in the up-down basis, which is the measure used by Jain \textit{et al}. \cite{Jain2021}. Note that, for all definitions used above, the non-trace preserving density matrix characteristic of recombining radical pair systems must be renormalized (by $\mathrm{Tr}[\hat{\rho}]$; which here simply yields the $k=0$ result). As the coherence measures $\mathcal{C}_{i}(t)$, $i \in \{ r, l_{1}, st \}$, are time-dependent we further introduce the coherence yield as the time-averaged coherence measure weighted by the exponential decay kinetics of the radical pair:
	\begin{align}
		\bar{\mathcal{C}}_{i} = k \int_{0}^{\infty}\mathcal{C}_{i}(t)\exp(-kt)\, \mathrm{d}t.
	\end{align}

	For comparison to the anisotropy, we further define measures accounting for the variation of $\bar{\mathcal{C}}_{i}$ with the magnetic field direction as
	\begin{align}
		\mu[\bar{\mathcal{C}}_{i}] &= \frac{ \bar{\mathcal{C}}_{i}(\hat{\rho}_{\max}) + \bar{\mathcal{C}}_{i}(\hat{\rho}_{\min})}{2}, \label{eq.mu_coherence}
	\end{align}
and 
	\begin{align}
		\Delta[\bar{\mathcal{C}}_{i}] &= \bar{\mathcal{C}}_{i}(\hat{\rho}_{\max})-\bar{\mathcal{C}}_{i}(\hat{\rho}_{\min}), \label{eq.delta_coherence}
	\end{align} 
	where $\hat{\rho}_{\max}$ and $\hat{\rho}_{\min}$ corresponds to the density matrix associated with maximum and minimum singlet yield over magnetic field orientations. Furthermore, a field-independent measure is introduced as $[\bar{\mathcal{C}}_{i}]_{B=0}$, which is evaluated for $B=0$.
	
	Cai and Plenio have suggested an alternative, operational global quantifier of coherence \cite{Cai2013} in the context of a chemical compass in terms of the field-independent ($B=0$) singlet recombination yield due to the coherent part of the initial density operator $\mathbb{GC}(\hat{\rho}(0))$, with $\mathbb{GC}(\hat{\rho}) = \hat{\rho} - \mathbb{IC}(\hat{\rho})$ evaluated in the eigenbasis of the hyperfine Hamiltonian:
	\begin{align}
		[\mathcal{C}_{y}^{\mathcal{G}}]_{B=0} = \Big\vert Y_{S}\Big(\hat{\rho}(0) = \mathrm{Tr}[\hat{P}_{S}]^{-1}\mathbb{GC}(\hat{P}_{S});B=0\Big)\Big\vert.
	\end{align}
	In this paper we utilise both this measure and a generalised field-dependent version (with the magnetic field in the extremal directions) to allow us to also evaluate $\mu[\mathcal{C}_{y}^{\mathcal{G}}]$ and $\Delta[\mathcal{C}_{y}^{\mathcal{G}}]$.
	
	\
	
	\noindent \textbf{Large spin system electronic coherence}. Previous studies on coherence in the avian compass typically use systems with a modest number of nuclear spins (only one or 5--6) and often randomly assumed hyperfine interactions \cite{Cai2013,Kominis2020,Gauger2011,Cai2012,Jain2021}. Under the assumptions as laid out above, we have been able to overcome this limitation and evaluate fidelities and coherence measures for flavin-tryptophan and flavin-tyrosine radical pairs with up to 21 nuclear spins and judiciously chosen hyperfine parameters, as summarized in the SI. Our implementation relied on GPU computing realized using CUDA, which vastly outperformed the corresponding calculations on CPUs. We have varied the relative orientation of the radicals, as parametrized by Euler angles, $\alpha$, $\beta$ and $\gamma$, defining the orientation of the flavin radical relative to the radical partner. Using an angular resolution of 3 degrees, we have probed 878,400 relative orientations for each combination of radicals. For each radical orientation, we sampled 10242 orientations of $\vartheta$ and $\varphi$ (5121 unique directions) to evaluate the fidelity measure $\Delta_{S}$ and several associated coherence measures as introduced above. In what follows, we focus on the compass anisotropy $\Delta_{S}$ and the coherence measures $\mu[\bar{\mathcal{C}}_{i}]$ and $\Delta[\bar{\mathcal{C}}_{i}]$, for both the ST and UD basis, and the $l_{1}$- and relative entropy of coherence quantifiers. Alternative measures are discussed in the SI, whereby qualitatively corresponding conclusions emerged.  For all simulations, a magnetic field strength of 50 $\mu$T (comparable to the geomagnetic field in Northern Europe) and a radical lifetime of $k^{-1}=1\mskip3mu\mu$s were assumed. The latter is in line with the order-of-magnitude lifetime of the magnetosensitive radical pairs in cryptochromes as observed for \textit{in vitro} studies \cite{Hore2016,Xu2021} and the anticipated spin relaxation times \cite{Kattnig2016a}. Results for larger magnetic field strength may be found in the SI; these show increased anisotropy but reduced correlation with coherence measures (Figs.\ S1 and S2).
	\begin{figure}[t]
	\centering
		\includegraphics[width=.9\linewidth]{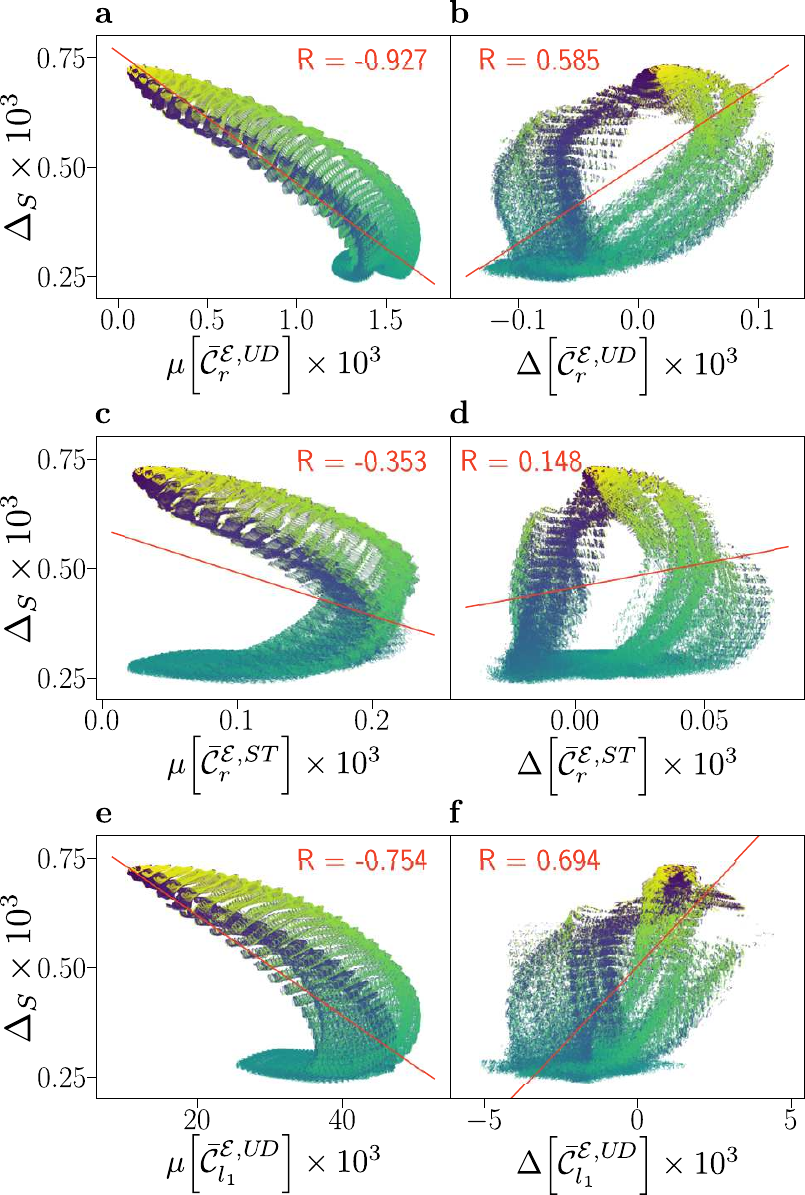}
		\caption{\label{fig:F_W} Compass sensitivity, as captured by the anisotropy $\Delta_{S}$, is plotted against electronic coherence measures $\mu[\bar{\mathcal{C}}_{i}]$ (\textbf{a}, \textbf{c}, \textbf{e}) and $\Delta[\bar{\mathcal{C}}_{i}]$ (\textbf{b}, \textbf{d}, \textbf{f}), for 878,400 relative orientations of flavin-tryptophan radical pairs. The anisotropy $\Delta_{S}$ quantifies the maximal change in the recombination quantum yield as the direction of the applied magnetic field is varied in the frame of the cryptochrome protein, with larger values corresponding to greater sensitivity. Here, we use the recombination weighted relative entropy of coherence $\bar{\mathcal{C}}_{r}$ defined in Eq.\ (\ref{eq.rel_entropy}), and the $l_{1}$-norm measure $\bar{\mathcal{C}}_{l_{1}}$ defined in Eq.\ (\ref{eq.l1_norm}) evaluated in the Zeeman basis, also known as the ``up-down'' (UD) basis, and the singlet-triplet (ST) basis. Coherence is assessed by its mean value $\mu[\bar{\mathcal{C}}_{i}]$ as defined in Eq.\ (\ref{eq.mu_coherence}), and difference $\Delta[\bar{\mathcal{C}}_{i}]$ as defined in Eq.\ (\ref{eq.delta_coherence}), with respect to the magnetic field directions associated with maximal and minimal recombination yield. Alternative measures are covered in the Supporting Information. Data has been coloured according to the relative orientation angle $\beta$, ranging from $0^{\circ}$ (blue) to $180^{\circ}$ (yellow) and a linear fit (red line) with associated Pearson correlation coefficient $R$ is displayed.}	
	\end{figure}

	For flavin-tryptophan systems the results are presented in Fig.\,\ref{fig:F_W}, which shows plots of the absolute anisotropy \textit{vs.}\ various electronic coherence measures. The absolute anisotropy $\Delta_{S}$ measures the maximal spread of the quantum yield of radical pair recombination as the direction of the geomagnetic field is varied relative to the sensory protein. Thus, $\Delta_{S}$ indicates the contrast of the sensor via a relative difference in reaction products with respect to magnetic field orientation and is widely accepted as a measure of compass sensitivity. In this way, Fig.\,\ref{fig:F_W} correlates this measure of compass sensitivity with measures quantifying the electronic coherence present during the lifetime of the radical pair. The data reveals diverse, i.e., not universal, and complex relations of fidelity and coherence measures. Surprisingly, an anti-correlation with respect to $\mu[\bar{\mathcal{C}}_{i}]$ is observed that is present in both the UD and ST basis, but larger in the former. Specifically, $\Delta_{S}$ is found to anti-correlate with $\mu[\bar{\mathcal{C}}_{r}^{\mathcal{E}, U\!D}]$, which is based on the same coherence measure as used by Jain \textit{et al}. \cite{Jain2021}, giving rise to a remarkable correlation coefficient of $R=-0.927$. In contrast, a positive (but weaker) correlation is observed with respect to the $\Delta[\bar{\mathcal{C}}_{r}^{\mathcal{E},UD}]$ measure. Similar results, shown in the SI, are found for correlations involving the relative anisotropy $\Gamma_{S}$ (Fig. S3), which is a measure of compass sensitivity $\Delta_{S}$ relative to the mean recombination yield.
	
	Furthermore, the results show structured subsets of the data that may present positive or negative correlations. In our data discussed below, and in comparable studies in the literature, such structures in the correlation data appear to be absent for a small number of nuclear spins. This suggests that for comprehensive considerations of coherence in biology, systems must be sufficiently and realistically complex.
	
	In the case of $\Delta[\bar{\mathcal{C}}_{r}^{\mathcal{E},ST}]$, the colouration of the plot according to the angle $\beta$ demonstrates a relationship to the bands in which small values of $\beta$ correspond to a stronger positive correlation (see Fig.\,\ref{fig:F_W}d). However, maximal anisotropy is realized for $\Delta[\bar{\mathcal{C}}_{r}^{\mathcal{E},ST}] \approx 0$, a condition which appears to likewise approximately hold for the other coherence anisotropy measures used (see Fig.\,\ref{fig:F_W}b and f). The implication is that, for this system, maximal compass sensitivity is not realized by minimizing electronic coherence in one extremal direction while maximizing it in the other, a strategy which, at least \textit{a priori}, might have appeared auspicious in maximizing yield differences and thus compass fidelity. 
	
\begin{figure}[t]
	\centering
		\includegraphics[width=.9\linewidth]{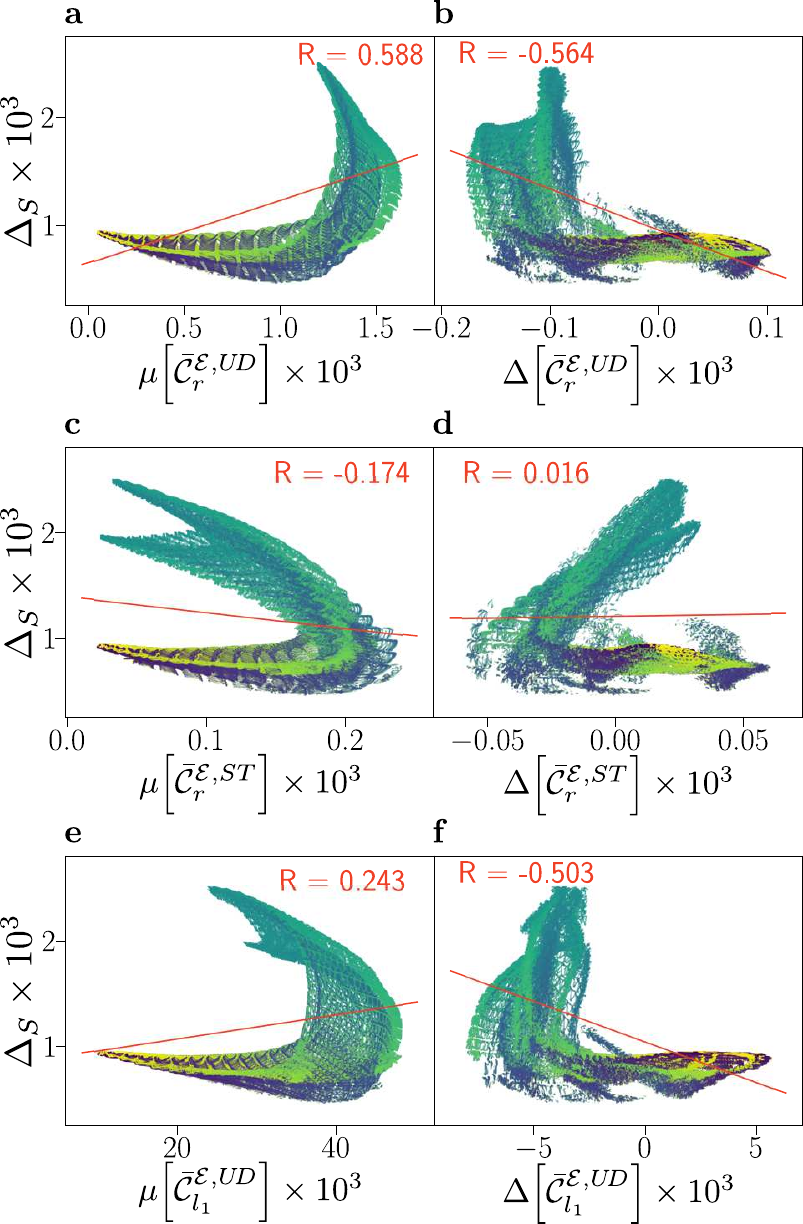}
		\caption{\label{fig:F_Y} Compass sensitivity, as captured by the anisotropy $\Delta_{S}$, is plotted against electronic coherence measures $\mu[\bar{\mathcal{C}}_{i}]$ (\textbf{a}, \textbf{c}, \textbf{e}) and $\Delta[\bar{\mathcal{C}}_{i}]$ (\textbf{b}, \textbf{d}, \textbf{f}), for 878,400 relative orientations of flavin-tyrosine radical pairs. Here, we use equivalent measures to Fig.\,\ref{fig:F_W}, where additional details are provided; alternative measures are detailed in the SI. Data has been coloured according to the relative orientation angle $\beta$, ranging from $0^{\circ}$ (blue) to $180^{\circ}$ (yellow). Both up-down (UD) and singlet-triplet (ST) bases are considered, and a linear fit (red line) with associated Pearson correlation coefficient $R$ is displayed.}	
	\end{figure}

	Results for flavin-tyrosine radical pairs are shown in Fig.\,\ref{fig:F_Y}, in which smaller correlations are found than
	for the flavin-tryptophan model. An anti-correlation is still observed for $\mu[\bar{\mathcal{C}}_{r}^{\mathcal{E},ST}]$ and $\Delta_{S}$, and likewise the weaker positive correlation persists between $\Delta[\bar{\mathcal{C}}_{r}^{\mathcal{E},ST}]$ and $\Delta_{S}$. As in the flavin-tryptophan models, 
	subsets of the data present positive and negative correlation. However, the structure of these subsets, and relationship to the angle $\beta$ is different.  Here, for mid-range values of $\beta$, which correspond to maximal compass sensitivity, the results suggest a marked 
	anti-correlation in $\mu[\bar{\mathcal{C}}_{r}^{\mathcal{E},ST}]$ and a strong positive correlation in $\Delta[\bar{\mathcal{C}}_{r}^{\mathcal{E},ST}]$. Similar results, shown in the SI (Figs.\ S3 and S4), are found for correlations involving  the relative anisotropy $\Gamma_{S}$.
	
	The results on large spin systems are unexpected in view of previous studies \cite{Hiscock2016,Kominis2020,Gauger2011,Fay2019} and thought provoking with respect to Jain \textit{et al}.’s suggestion of a compass operating in a regime of low coherence \cite{Jain2021}. Here, our results based on the same coherence measure as used Jain \textit{et al}. ($\bar{\mathcal{C}}_{r}^{\mathcal{E}, U\!D}$), appear to not only corroborate this conclusion, but lead to the supposition that the lack of electronic coherence is advantageous. This could imply an incoherent, relaxation-driven \cite{Tiersch2012,Carrillo2015} character of the avian compass.
	
	\
	
	\noindent \textbf{Small spin system electronic and global coherence}. We note that the coherence measures employed above, including $\mathcal{C}_{r}^{\mathcal{E},U\!D}$ as used by Jain \textit{et al}., have not previously been assessed with respect to their principal predictive power of compass performance. We have thus systematically explored the correlation of compass fidelity and the various coherence measures employed, both for systems of reference-probe topology \cite{Lee2014,Procopio2020} (with all hyperfine interactions confined to one radical) and systems with more symmetrically distributed hyperfine interactions, and for radical pair lifetimes of $k^{-1}=1\mskip3mu\mu$s and $k^{-1}=10\mskip3mu\mu$s. Details of this exploration are provided in the SI (Figs. S5–S8). In brief, we again find that there is significant variation of correlations for different systems and measures, overall supporting the view that electronic coherence is not the most effective predictor of compass sensitivity. In particular systems that combine the hyperfine interactions of N5 and N10 in flavin with random hyperfine couplings in the other radical exhibit low correlation between anisotropy and electronic coherence. However, as was the case in the study presented by Cai and Plenio \cite{Cai2013}, a stronger correlation is found with global coherence measures, in particular $\mathcal{C}^{\mathcal{G}}_{y}$. This measure corresponds to the coherent contribution to the singlet yield, and its broad success suggests that it is important that the global coherence quantifier relates to operation. Further measures and correlation plots supporting these findings can be found in the SI (Figs. S5–S8).
	
	\
	
	\noindent \textbf{Large spin system global coherence}. The confirmed applicability of $[\mathcal{C}_{y}^{\mathcal{G}}]_{B=0}$ motivated us to attempt a reassessment of the coherence-fidelity correlation of the large spin systems based on this global coherence measure. While global coherence measures are in general prohibitively expensive to evaluate for spin systems of the relevant size, $[\mathcal{C}_{y}^{\mathcal{G}}]_{B=0}$ can be approximated with reasonable effort from the analytical expression for the singlet yield (eq.\ (S14) in the SI; implemented for graphics processing unit using CUDA). Using Monte Carlo sampling of the matrix elements of the singlet projection operator in the eigenbasis of the combined Hilbert space of both radicals assists in accelerating the calculation (see SI). Utilising this we have been able to evaluate $[\mathcal{C}_{y}^{\mathcal{G}}]_{B=0}$ of the flavin-tryptophan and flavin-tyrosine systems for a representative number of relative orientations. 

	\begin{figure}[t]
	\centering
	\includegraphics[width=\linewidth]{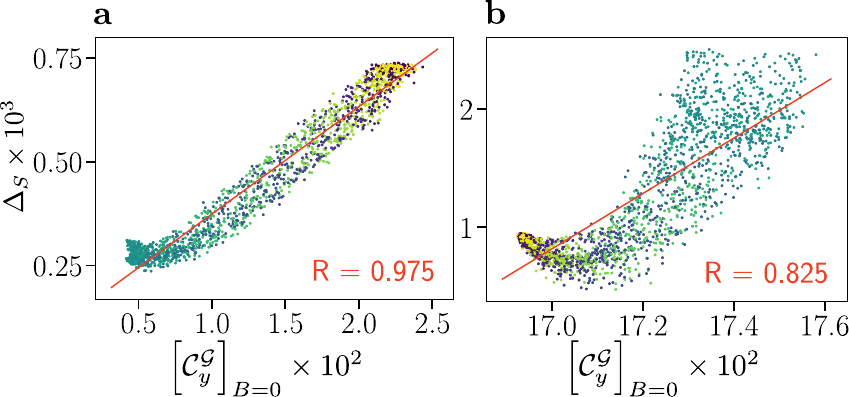}
	\caption{\label{fig:global_coherence}Compass anisotropy $\Delta_{S}$ as a measure of compass sensitivity is plotted against the global measure of coherence $[\mathcal{C}_{y}^{\mathcal{G}}]_{B=0}$, an operational measure that represents the coherence contribution of the total nuclear and electronic state to the singlet yield evaluated in the absence of a magnetic field. \textbf{a} Correlation plot for the flavin-tryptophan model. \textbf{b} Correlation plot for flavin-tyrosine model.  Data has been coloured according to the radical relative orientation angle $\beta$, ranging from $0^{\circ}$ (blue) to $180^{\circ}$ (yellow), and a linear fit (red line) with associated Pearson correlation coefficient $R$ is displayed.}	
	\end{figure}

	In Fig.\ \ref{fig:global_coherence} we plot compass anisotropy $\Delta_{S}$ of these systems as a function of global coherence $[\mathcal{C}_{y}^{\mathcal{G}}]_{B=0}$. For both systems, a strong positive correlation of $[\mathcal{C}_{y}^{\mathcal{G}}]_{B=0}$ and $\Delta_{S}$ is found, thereby reinforcing the results for 5-6 nuclear spins, as studied previously \cite{Cai2013} and above. Specifically, we find a remarkable correlation for flavin-tryptophan for which $R = +0.975$ based on 1784 randomly chosen relative orientations, in stark contrast to the anti-correlation of $R = -0.927$ found with the electronic coherence measure $\mu[\bar{\mathcal{C}}_{r}^{\mathcal{E},U\!D}]$. The results for flavin-tyrosine are more structured, but likewise give rise to a strong overall correlation with $R = +0.825$. This result confirms the supposition of Cai and Plenio, who proposed that global system-environment coherence is crucial for the compass sensitivity, based on simpler model systems \cite{Cai2013}. Alongside our other results, 
	this implies that the total system cannot be reduced to consideration of electronic subsystems. Nuclear coherences and nuclear-electronic coherences are an integral part of the processes. 
	
	\section*{Discussion}
	\noindent The radical-pair compass, by its design, relies on the coherent interconversion of singlet and triplet electronic states and, thus, the presence of a certain degree of singlet-triplet coherence is essential \cite{Hore2016,Le2020,Tiersch2012}. However, we find that the amount of electronic coherence does not generally correlate with increased compass performance across differently oriented systems. Instead, in both large spin systems considered, we observe a surprising anti-correlation, between selected mean electronic coherence measures and the reaction yield anisotropy. In contrast, global coherence measures are found to predict compass performance more effectively than electronic measures. 
	
	\
	
	\noindent \textbf{Quantum biology and coherence.} Overall, our results suggest that nature, rather than maximising and prolonging the possible electronic coherence, must utilise an interplay of several factors on relevant timescales to increase compass sensitivity. These include the effect of the magnetic field on coherent dynamics, engineering of radical pair lifetimes, decay channels and the hyperfine-driven singlet-triplet dissipation \cite{Kattnig2016a,Kattnig2016b,Tiersch2012}. Complex radical systems cannot be reduced to consideration of electronic subsystems. Nuclear coherences and nuclear-electronic coherences are an integral part of the processes, which must not be neglected. On the contrary, for FAD$^{\bullet-}$/W$^{\bullet+}$ the anti-correlated electronic coherence measures suggest that global coherence is realized by sacrificing electronic coherence to boost the compass fidelity. The crucial relevance of the nuclear degrees of freedom has also become apparent in a recent study exploring the effects of accumulated nuclear polarisation for repeatedly re-excited radical pairs as a means to boost anisotropic MFEs \cite{Wong2021}. Conversely, incoherent operations on nuclear spins can have a signiﬁcant effect on the singlet-triplet coherence \cite{Kominis2020}.
	
	\
	
	\noindent \textbf{Quantum advantage at short lifetimes.} Non-classicality in complex spin systems representative of flavin-tryptophan has previously been postulated based on the comparison with semi-classical calculations \cite{Fay2019,Manolopoulos2013,Lewis2014}. For long coherent lifetimes, an increase in sensitivity emerging as a sharp spike in the fractional yield of the singlet product \cite{Hiscock2016} is apparent in quantum simulations, but absent from the semi-classical description, thereby demonstrating a clear-cut quantum advantage. This ``quantum needle'', however, only emerges significantly for coherent lifetimes that are likely unrealistically long, at least in view of the expected spin relaxation rates in cryptochromes and direct animal behavioural experiments using rf-magnetic fields to interfere with compass navigation (specifically, their frequency dependence) \cite{Kobylkov2019,Kattnig2016b,Kattnig2016a}. In this regard, the correlation of compass fidelity and the coherence measure $[\mathcal{C}_{y}^{\mathcal{G}}]_{B=0}$ as found here is remarkable, because it suggests a manifest quantum advantage despite the comparably short lifetime of $1\mskip3mu\mu$s, for which the semi-classical and fully quantum approaches concur. 
	
	\
	
	\noindent \textbf{Realistic model complexity.} Advancing our understanding of non-trivial quantum effects in biology requires the precise characterisation of putative quantum feats as well as asserting whether this non-classicality enables robust function. In the context of the avian compass, resourceful non-classicality has been implicated based on the correlation of compass fidelity and coherence measures. However, previous studies have focused on toy models, which did not live up to the complexity expected for the actual system. Therefore, at best the principal feasibility could be deduced, while the physiological relevance remained opaque. Here, a robust correlation of compass fidelity and global coherence measures is established for large spin systems and realistic parameters, thereby corroborating the view that coherence---but not electronic coherence---is truly a quantum resource for magnetoreception.

	In summary, our results demonstrate Cai and Plenio’s claim \cite{Cai2013}, that global coherence is a resource, in the complex system regime, whilst simultaneously supporting the claim \cite{Jain2021} of Jain \textit{et al.}\ that sensitivity can be obtained without sustained electronic coherence for some systems. The connection between seemingly contrasting results, concerning electronic coherence as a resource, is found when it is considered in the wider setting of several diverse systems and complexities as presented in our study. Future studies could steadily increase the relevant complexity of models in a tractable manner and consider alternative models of the radical reaction \cite{Keens2018,Kattnig2017} such as the flavin semiquinone/O$_2^{\bullet -}$ radical pair, studied e.g. as a function of the binding site of the latter. As O$_2^{\bullet-}$ is devoid of hyperfine interactions, this system is invariant to the mutual orientation of the radicals and thus not amenable to the study design as chosen here. It should be noted that even the large systems studied here have been (necessarily) idealized. The asymmetry of the reactions has been neglected, by setting $k=k_{S}=k_{T}$, and inter-radical interactions (such as electron-electron dipolar interactions) has been ignored, both of which have been associated with peculiar quantum feats \cite{Dellis2012,Pedersen2016,Kattnig2017,Keens2018,Babcock2020,Babcock2021}. Additionally, the open quantum system nature of the system could be analysed in further detail \cite{Adams2018, Berkelbach2020}, employing numerically exact methods \cite{Tanimura2020,Suess2014,Hartmann2017} for greater accuracy and to identify if non-Markovian dynamics \cite{Breuer2015,DeVega2017,Wu2020} are of importance in realistic natural and artificial systems. We anticipate that the approach and observations of this paper will aid these future endeavours through identification of the most effective use of coherence measures, and by paving a way to a more realistic assessment of the putatively utilitarian character of coherence in magnetoreception. Regarding this, we recommend that results should always be rationalised with respect to the complexity of the system.

\section*{Methods}
\noindent \textbf{Numerical implementation.} We have considered radical pair systems with up to 21 nuclear spins. The ability to simulate such large spin systems relies on the separability of the problem in the absence of inter-radical interactions, which allows us to express the electronic density operator in terms of spin correlation tensors (SCTs), defined for each individual radical \cite{Atkins2019,Manolopoulos2013}. Thus the problem of treating 21 nuclear spins for flavin/tryptophan reduces to independently propagating spin systems comprising 12 nuclear spins for flavin, and 9 nuclear spins for tryptophan. To assess the effect of the mutual orientation of the two radicals in space we additionally rely on an implementation that rotates the SCTs rather than recalculating them \cite{Atkins2019}. The SCTs have been calculated on graphic processing units (GPU) using purpose-designed CUDA kernels, which vastly outperforms similar CPU computations. Taken together this provides a means to calculate electronic coherence quantifiers and singlet recombination yields of complex radical systems subject to arbitrary relative reorientations of the constituent radicals. We have also devised a GPU implementation to calculate recombination yields directly, which we used alongside Monte Carlo sampling of the expectation values of the singlet projection operator in the eigenbasis of the Hamiltonian to calculate the global coherence measure $[\mathcal{C}_{y}^{\mathcal{G}}]_{B=0}$. 

Additional details on the implementation, including the CUDA-kernels devised, are provided in the Supporting Information.

\section*{Data availability}
The datasets generated and analyzed during the current study have been made available via Open Research Exeter (ORE), the University of Exeter’s online repository: \url{https://doi.org/10.24378/exe.3923}.

\section*{Acknowledgments}
\noindent We gladly acknowledge the use of the University of Exeter High-Performance Computing facility. This work was supported by the UK Defence Science and Technology Laboratory (DSTLX-1000139168), the Office of Naval Research (ONR award number N62909-21-1-2018) and the EPSRC (grant EP/V047175/1). The robin in Fig.\ 1 is extracted from the artwork ``Robin Vintage Art Poster'' released under Public Domain license by Karen Arnold.

\section*{References}
\bibliographystyle{naturemag}
\bibliography{spincoherenceV2.bib}

\begin{thebibliography}{10}
\expandafter\ifx\csname url\endcsname\relax
  \def\url#1{\texttt{#1}}\fi
\expandafter\ifx\csname urlprefix\endcsname\relax\def\urlprefix{URL }\fi
\providecommand{\bibinfo}[2]{#2}
\providecommand{\eprint}[2][]{\url{#2}}

\bibitem{Ball2011}
\bibinfo{author}{Ball, P.}
\newblock \bibinfo{title}{{Physics of life: The dawn of quantum biology}}.
\newblock \emph{\bibinfo{journal}{Nature}} \textbf{\bibinfo{volume}{474}},
  \bibinfo{pages}{272--274} (\bibinfo{year}{2011}).
\newblock \urlprefix\url{https://www.nature.com/articles/474272a}.

\bibitem{Lambert2013}
\bibinfo{author}{Lambert, N.} \emph{et~al.}
\newblock \bibinfo{title}{{Quantum biology}}.
\newblock \emph{\bibinfo{journal}{Nature Physics}}
  \textbf{\bibinfo{volume}{9}}, \bibinfo{pages}{10--18} (\bibinfo{year}{2013}).
\newblock \urlprefix\url{https://www.nature.com/articles/nphys2474}.

\bibitem{McFadden2018}
\bibinfo{author}{McFadden, J.} \& \bibinfo{author}{Al-Khalili, J.}
\newblock \bibinfo{title}{{The origins of quantum biology}}.
\newblock \emph{\bibinfo{journal}{Proceedings of the Royal Society A}}
  \textbf{\bibinfo{volume}{474}}, \bibinfo{pages}{20180674}
  (\bibinfo{year}{2018}).
\newblock \urlprefix\url{http://dx.doi.org/10.1098/rspa.2018.0674}.

\bibitem{Marais2018}
\bibinfo{author}{Marais, A.} \emph{et~al.}
\newblock \bibinfo{title}{{The future of quantum biology}}.
\newblock \emph{\bibinfo{journal}{Journal of the Royal Society Interface}}
  \textbf{\bibinfo{volume}{15}}, \bibinfo{pages}{20180640}
  (\bibinfo{year}{2018}).
\newblock \urlprefix\url{https://doi.org/10.1098/rsif.2018.0640}.

\bibitem{Kim2021}
\bibinfo{author}{Kim, Y.} \emph{et~al.}
\newblock \bibinfo{title}{{Quantum biology: An update and perspective}}.
\newblock \emph{\bibinfo{journal}{Quantum Reports}}
  \textbf{\bibinfo{volume}{3}}, \bibinfo{pages}{80--126}
  (\bibinfo{year}{2021}).
\newblock \urlprefix\url{https://www.mdpi.com/2624-960X/3/1/6}.

\bibitem{Romero2014}
\bibinfo{author}{Romero, E.} \emph{et~al.}
\newblock \bibinfo{title}{{Quantum coherence in photosynthesis for efficient
  solar-energy conversion}}.
\newblock \emph{\bibinfo{journal}{Nature Physics}}
  \textbf{\bibinfo{volume}{10}}, \bibinfo{pages}{676--682}
  (\bibinfo{year}{2014}).
\newblock \urlprefix\url{https://www.nature.com/articles/nphys3017}.

\bibitem{Kominis2015}
\bibinfo{author}{Kominis, I.~K.}
\newblock \bibinfo{title}{{The radical-pair mechanism as a paradigm for the
  emerging science of quantum biology}}.
\newblock \emph{\bibinfo{journal}{Modern Physics Letters B}}
  \textbf{\bibinfo{volume}{29}}, \bibinfo{pages}{1530013}
  (\bibinfo{year}{2015}).
\newblock \urlprefix\url{https://doi.org/10.1142/S0217984915300136}.

\bibitem{Duan2017}
\bibinfo{author}{Duan, H.-G.} \emph{et~al.}
\newblock \bibinfo{title}{{Nature does not rely on long-lived electronic
  quantum coherence for photosynthetic energy transfer}}.
\newblock \emph{\bibinfo{journal}{Proceedings of the National Academy of
  Sciences of the United States of America}} \textbf{\bibinfo{volume}{114}},
  \bibinfo{pages}{8493--8498} (\bibinfo{year}{2017}).
\newblock \urlprefix\url{https://www.pnas.org/doi/10.1073/pnas.1702261114}.

\bibitem{Thyrhaug2018}
\bibinfo{author}{Thyrhaug, E.} \emph{et~al.}
\newblock \bibinfo{title}{{Identification and characterization of diverse
  coherences in the Fenna–Matthews–Olson complex}}.
\newblock \emph{\bibinfo{journal}{Nature Chemistry}}
  \textbf{\bibinfo{volume}{10}}, \bibinfo{pages}{780--786}
  (\bibinfo{year}{2018}).
\newblock \urlprefix\url{https://www.nature.com/articles/s41557-018-0060-5}.

\bibitem{Cao2020}
\bibinfo{author}{Cao, J.} \emph{et~al.}
\newblock \bibinfo{title}{{Quantum biology revisited}}.
\newblock \emph{\bibinfo{journal}{Science Advances}}
  \textbf{\bibinfo{volume}{6}}, \bibinfo{pages}{eaaz4888}
  (\bibinfo{year}{2020}).
\newblock \urlprefix\url{https://www.science.org/doi/10.1126/sciadv.aaz4888}.

\bibitem{Baumgratz2014}
\bibinfo{author}{Baumgratz, T.}, \bibinfo{author}{Cramer, M.} \&
  \bibinfo{author}{Plenio, M.~B.}
\newblock \bibinfo{title}{{Quantifying coherence}}.
\newblock \emph{\bibinfo{journal}{Physical Review Letters}}
  \textbf{\bibinfo{volume}{113}}, \bibinfo{pages}{140401}
  (\bibinfo{year}{2014}).
\newblock
  \urlprefix\url{https://journals.aps.org/prl/abstract/10.1103/PhysRevLett.113.140401}.

\bibitem{Streltsov2016}
\bibinfo{author}{Streltsov, A.}, \bibinfo{author}{Adesso, G.} \&
  \bibinfo{author}{Plenio, M.~B.}
\newblock \bibinfo{title}{{Colloquium: Quantum coherence as a resource}}.
\newblock \emph{\bibinfo{journal}{Reviews of Modern Physics}}
  \textbf{\bibinfo{volume}{89}}, \bibinfo{pages}{041003}
  (\bibinfo{year}{2017}).
\newblock
  \urlprefix\url{https://journals.aps.org/rmp/abstract/10.1103/RevModPhys.89.041003}.

\bibitem{Winter2016}
\bibinfo{author}{Winter, A.} \& \bibinfo{author}{Yang, D.}
\newblock \bibinfo{title}{{Operational resource theory of coherence}}.
\newblock \emph{\bibinfo{journal}{Physical Review Letters}}
  \textbf{\bibinfo{volume}{116}}, \bibinfo{pages}{120404}
  (\bibinfo{year}{2016}).
\newblock
  \urlprefix\url{https://journals.aps.org/prl/abstract/10.1103/PhysRevLett.116.120404}.

\bibitem{Hore2016}
\bibinfo{author}{Hore, P.~J.} \& \bibinfo{author}{Mouritsen, H.}
\newblock \bibinfo{title}{{The radical-pair mechanism of magnetoreception}}.
\newblock \emph{\bibinfo{journal}{Annual Review of Biophysics}}
  \textbf{\bibinfo{volume}{45}}, \bibinfo{pages}{299--344}
  (\bibinfo{year}{2016}).
\newblock \urlprefix\url{https://pubmed.ncbi.nlm.nih.gov/27216936/}.

\bibitem{Ritz2000}
\bibinfo{author}{Ritz, T.}, \bibinfo{author}{Adem, S.} \&
  \bibinfo{author}{Schulten, K.}
\newblock \bibinfo{title}{{A model for photoreceptor-based magnetoreception in
  birds}}.
\newblock \emph{\bibinfo{journal}{Biophysical Journal}}
  \textbf{\bibinfo{volume}{78}}, \bibinfo{pages}{707--718}
  (\bibinfo{year}{2000}).
\newblock \urlprefix\url{https://doi.org/10.1016/S0006-3495(00)76629-X}.

\bibitem{Tiersch2012}
\bibinfo{author}{Tiersch, M.} \& \bibinfo{author}{Briegel, H.~J.}
\newblock \bibinfo{title}{{Decoherence in the chemical compass: the role of
  decoherence for avian magnetoreception}}.
\newblock \emph{\bibinfo{journal}{Philosophical Transactions of the Royal
  Society A}} \textbf{\bibinfo{volume}{370}}, \bibinfo{pages}{4517--4540}
  (\bibinfo{year}{2012}).
\newblock
  \urlprefix\url{https://royalsocietypublishing.org/doi/abs/10.1098/rsta.2011.0488}.

\bibitem{Hiscock2016}
\bibinfo{author}{Hiscock, H.~G.} \emph{et~al.}
\newblock \bibinfo{title}{{The quantum needle of the avian magnetic compass}}.
\newblock \emph{\bibinfo{journal}{Proceedings of the National Academy of
  Sciences of the United States of America}} \textbf{\bibinfo{volume}{113}},
  \bibinfo{pages}{4634--4639} (\bibinfo{year}{2016}).
\newblock \urlprefix\url{www.pnas.org/cgi/doi/10.1073/pnas.1600341113}.

\bibitem{Kobylkov2019}
\bibinfo{author}{Kobylkov, D.} \emph{et~al.}
\newblock \bibinfo{title}{{Electromagnetic 0.1–100 kHz noise does not disrupt
  orientation in a night-migrating songbird implying a spin coherence lifetime
  of less than 10 µs}}.
\newblock \emph{\bibinfo{journal}{Journal of the Royal Society Interface}}
  \textbf{\bibinfo{volume}{16}}, \bibinfo{pages}{20190716}
  (\bibinfo{year}{2019}).
\newblock
  \urlprefix\url{https://royalsocietypublishing.org/doi/abs/10.1098/rsif.2019.0716}.

\bibitem{Kattnig2016b}
\bibinfo{author}{Kattnig, D.~R.}, \bibinfo{author}{Sowa, J.~K.},
  \bibinfo{author}{Solov'Yov, I.~A.} \& \bibinfo{author}{Hore, P.~J.}
\newblock \bibinfo{title}{{Electron spin relaxation can enhance the performance
  of a cryptochrome-based magnetic compass sensor}}.
\newblock \emph{\bibinfo{journal}{New Journal of Physics}}
  \textbf{\bibinfo{volume}{18}}, \bibinfo{pages}{063007}
  (\bibinfo{year}{2016}).
\newblock
  \urlprefix\url{https://iopscience.iop.org/article/10.1088/1367-2630/18/6/063007}.

\bibitem{Kattnig2016a}
\bibinfo{author}{Kattnig, D.~R.}, \bibinfo{author}{Solov'yov, I.~A.} \&
  \bibinfo{author}{Hore, P.~J.}
\newblock \bibinfo{title}{{Electron spin relaxation in cryptochrome-based
  magnetoreception}}.
\newblock \emph{\bibinfo{journal}{Physical Chemistry Chemical Physics}}
  \textbf{\bibinfo{volume}{18}}, \bibinfo{pages}{12443--12456}
  (\bibinfo{year}{2016}).

\bibitem{Xu2021}
\bibinfo{author}{Xu, J.} \emph{et~al.}
\newblock \bibinfo{title}{{Magnetic sensitivity of cryptochrome 4 from a
  migratory songbird}}.
\newblock \emph{\bibinfo{journal}{Nature}} \textbf{\bibinfo{volume}{594}},
  \bibinfo{pages}{535--540} (\bibinfo{year}{2021}).
\newblock \urlprefix\url{https://www.nature.com/articles/s41586-021-03618-9}.

\bibitem{P2011}
\bibinfo{author}{P, M.} \& \bibinfo{author}{M, A.}
\newblock \bibinfo{title}{{Light-activated cryptochrome reacts with molecular
  oxygen to form a flavin-superoxide radical pair consistent with
  magnetoreception}}.
\newblock \emph{\bibinfo{journal}{The Journal of Biological Chemistry}}
  \textbf{\bibinfo{volume}{286}}, \bibinfo{pages}{21033--21040}
  (\bibinfo{year}{2011}).
\newblock \urlprefix\url{https://pubmed.ncbi.nlm.nih.gov/21467031/}.

\bibitem{Wiltschko2016}
\bibinfo{author}{Wiltschko, R.}, \bibinfo{author}{Ahmad, M.},
  \bibinfo{author}{Nie{\ss}ner, C.}, \bibinfo{author}{Gehring, D.} \&
  \bibinfo{author}{Wiltschko, W.}
\newblock \bibinfo{title}{{Light-dependent magnetoreception in birds: the
  crucial step occurs in the dark}}.
\newblock \emph{\bibinfo{journal}{Journal of The Royal Society Interface}}
  \textbf{\bibinfo{volume}{13}}, \bibinfo{pages}{20151010}
  (\bibinfo{year}{2016}).
\newblock
  \urlprefix\url{https://royalsocietypublishing.org/doi/abs/10.1098/rsif.2015.1010}.

\bibitem{Player2019}
\bibinfo{author}{Player, T.~C.} \& \bibinfo{author}{Hore, P.~J.}
\newblock \bibinfo{title}{{Viability of superoxide-containing radical pairs as
  magnetoreceptors}}.
\newblock \emph{\bibinfo{journal}{The Journal of Chemical Physics}}
  \textbf{\bibinfo{volume}{151}}, \bibinfo{pages}{225101}
  (\bibinfo{year}{2019}).
\newblock \urlprefix\url{https://aip.scitation.org/doi/abs/10.1063/1.5129608}.

\bibitem{Atkins2019}
\bibinfo{author}{Atkins, C.}, \bibinfo{author}{Bajpai, K.},
  \bibinfo{author}{Rumball, J.} \& \bibinfo{author}{Kattnig, D.~R.}
\newblock \bibinfo{title}{{On the optimal relative orientation of radicals in
  the cryptochrome magnetic compass}}.
\newblock \emph{\bibinfo{journal}{The Journal of Chemical Physics}}
  \textbf{\bibinfo{volume}{151}}, \bibinfo{pages}{065103}
  (\bibinfo{year}{2019}).
\newblock \urlprefix\url{https://aip.scitation.org/doi/abs/10.1063/1.5115445}.

\bibitem{Lee2014}
\bibinfo{author}{Lee, A.~A.} \emph{et~al.}
\newblock \bibinfo{title}{{Alternative radical pairs for cryptochrome-based
  magnetoreception}}.
\newblock \emph{\bibinfo{journal}{Journal of The Royal Society Interface}}
  \textbf{\bibinfo{volume}{11}}, \bibinfo{pages}{20131063}
  (\bibinfo{year}{2014}).
\newblock
  \urlprefix\url{https://royalsocietypublishing.org/doi/abs/10.1098/rsif.2013.1063}.

\bibitem{Procopio2020}
\bibinfo{author}{Procopio, M.} \& \bibinfo{author}{Ritz, T.}
\newblock \bibinfo{title}{{The reference-probe model for a robust and optimal
  radical-pair-based magnetic compass sensor}}.
\newblock \emph{\bibinfo{journal}{The Journal of Chemical Physics}}
  \textbf{\bibinfo{volume}{152}}, \bibinfo{pages}{065104}
  (\bibinfo{year}{2020}).
\newblock \urlprefix\url{https://aip.scitation.org/doi/abs/10.1063/1.5128128}.

\bibitem{Cai2013}
\bibinfo{author}{Cai, J.} \& \bibinfo{author}{Plenio, M.~B.}
\newblock \bibinfo{title}{{Chemical compass model for avian magnetoreception as
  a quantum coherent device}}.
\newblock \emph{\bibinfo{journal}{Physical Review Letters}}
  \textbf{\bibinfo{volume}{111}}, \bibinfo{pages}{230503}
  (\bibinfo{year}{2013}).
\newblock
  \urlprefix\url{https://link.aps.org/doi/10.1103/PhysRevLett.111.230503}.

\bibitem{Kominis2020}
\bibinfo{author}{Kominis, I.~K.}
\newblock \bibinfo{title}{{Quantum relative entropy shows singlet-triplet
  coherence is a resource in the radical-pair mechanism of biological magnetic
  sensing}}.
\newblock \emph{\bibinfo{journal}{Physical Review Research}}
  \textbf{\bibinfo{volume}{2}}, \bibinfo{pages}{023206} (\bibinfo{year}{2020}).
\newblock
  \urlprefix\url{https://journals.aps.org/prresearch/abstract/10.1103/PhysRevResearch.2.023206}.

\bibitem{Hogben2012}
\bibinfo{author}{Hogben, H.~J.}, \bibinfo{author}{Biskup, T.} \&
  \bibinfo{author}{Hore, P.~J.}
\newblock \bibinfo{title}{{Entanglement and sources of magnetic anisotropy in
  radical pair-based avian magnetoreceptors}}.
\newblock \emph{\bibinfo{journal}{Physical Review Letters}}
  \textbf{\bibinfo{volume}{109}}, \bibinfo{pages}{220501}
  (\bibinfo{year}{2012}).
\newblock
  \urlprefix\url{https://journals.aps.org/prl/abstract/10.1103/PhysRevLett.109.220501}.

\bibitem{Le2020}
\bibinfo{author}{Le, T.~P.} \& \bibinfo{author}{Olaya-Castro, A.}
\newblock \bibinfo{title}{{Basis-independent system-environment coherence is
  necessary to detect magnetic field direction in an avian-inspired quantum
  magnetic sensor}} (\bibinfo{year}{2020}).
\newblock \urlprefix\url{https://arxiv.org/abs/2011.15016v1}.
\newblock \eprint{2011.15016}.

\bibitem{Jain2021}
\bibinfo{author}{Jain, R.}, \bibinfo{author}{Poonia, V.~S.},
  \bibinfo{author}{Saha, K.}, \bibinfo{author}{Saha, D.} \&
  \bibinfo{author}{Ganguly, S.}
\newblock \bibinfo{title}{{The avian compass can be sensitive even without
  sustained electron spin coherence}}.
\newblock \emph{\bibinfo{journal}{Proceedings of the Royal Society A}}
  \textbf{\bibinfo{volume}{477}}, \bibinfo{pages}{20200778}
  (\bibinfo{year}{2021}).
\newblock
  \urlprefix\url{https://royalsocietypublishing.org/doi/abs/10.1098/rspa.2020.0778}.

\bibitem{Procopio2016}
\bibinfo{author}{Procopio, M.} \& \bibinfo{author}{Ritz, T.}
\newblock \bibinfo{title}{{Inhomogeneous ensembles of radical pairs in chemical
  compasses}}.
\newblock \emph{\bibinfo{journal}{Scientific Reports 2016 6:1}}
  \textbf{\bibinfo{volume}{6}}, \bibinfo{pages}{1--17} (\bibinfo{year}{2016}).
\newblock \urlprefix\url{https://www.nature.com/articles/srep35443}.

\bibitem{Haberkorn1976}
\bibinfo{author}{Haberkorn, R.}
\newblock \bibinfo{title}{{Density matrix description of spin-selective radical
  pair reactions}}.
\newblock \emph{\bibinfo{journal}{Molecular Physics}}
  \textbf{\bibinfo{volume}{32}}, \bibinfo{pages}{1491--1493}
  (\bibinfo{year}{1976}).
\newblock
  \urlprefix\url{https://www.tandfonline.com/doi/abs/10.1080/00268977600102851}.

\bibitem{Fay2018}
\bibinfo{author}{Fay, T.~P.}, \bibinfo{author}{Lindoy, L.~P.} \&
  \bibinfo{author}{Manolopoulos, D.~E.}
\newblock \bibinfo{title}{{Spin-selective electron transfer reactions of
  radical pairs: Beyond the Haberkorn master equation}}.
\newblock \emph{\bibinfo{journal}{The Journal of Chemical Physics}}
  \textbf{\bibinfo{volume}{149}}, \bibinfo{pages}{064107}
  (\bibinfo{year}{2018}).
\newblock \urlprefix\url{https://aip.scitation.org/doi/abs/10.1063/1.5041520}.

\bibitem{Manolopoulos2013}
\bibinfo{author}{Manolopoulos, D.~E.} \& \bibinfo{author}{Hore, P.~J.}
\newblock \bibinfo{title}{{An improved semiclassical theory of radical pair
  recombination reactions}}.
\newblock \emph{\bibinfo{journal}{The Journal of Chemical Physics}}
  \textbf{\bibinfo{volume}{139}}, \bibinfo{pages}{124106}
  (\bibinfo{year}{2013}).
\newblock \urlprefix\url{https://aip.scitation.org/doi/abs/10.1063/1.4821817}.

\bibitem{Schulten1978}
\bibinfo{author}{Schulten, K.} \& \bibinfo{author}{Wolynes, P.~G.}
\newblock \bibinfo{title}{{Semiclassical description of electron spin motion in
  radicals including the effect of electron hopping}}.
\newblock \emph{\bibinfo{journal}{The Journal of Chemical Physics}}
  \textbf{\bibinfo{volume}{68}}, \bibinfo{pages}{3292--3297}
  (\bibinfo{year}{1978}).
\newblock \urlprefix\url{https://doi.org/10.1063/1.436135}.

\bibitem{Gauger2011}
\bibinfo{author}{Gauger, E.~M.}, \bibinfo{author}{Rieper, E.},
  \bibinfo{author}{Morton, J. J.~L.}, \bibinfo{author}{Benjamin, S.~C.} \&
  \bibinfo{author}{Vedral, V.}
\newblock \bibinfo{title}{{Sustained quantum coherence and entanglement in the
  avian compass}}.
\newblock \emph{\bibinfo{journal}{Physical Review Letters}}
  \textbf{\bibinfo{volume}{106}}, \bibinfo{pages}{040503}
  (\bibinfo{year}{2011}).
\newblock
  \urlprefix\url{https://journals.aps.org/prl/abstract/10.1103/PhysRevLett.106.040503}.

\bibitem{Pauls2011}
\bibinfo{author}{Pauls, J.~A.}, \bibinfo{author}{Zhang, Y.},
  \bibinfo{author}{Berman, G.~P.} \& \bibinfo{author}{Kais, S.}
\newblock \bibinfo{title}{{Quantum coherence and entanglement in the avian
  compass}}.
\newblock \emph{\bibinfo{journal}{Physical Review E}}
  \textbf{\bibinfo{volume}{87}}, \bibinfo{pages}{062704}
  (\bibinfo{year}{2011}).
\newblock
  \urlprefix\url{https://journals.aps.org/pre/abstract/10.1103/PhysRevE.87.062704}.

\bibitem{Kritsotakis2014}
\bibinfo{author}{Kritsotakis, M.} \& \bibinfo{author}{Kominis, I.~K.}
\newblock \bibinfo{title}{{Retrodictive derivation of the radical-ion-pair
  master equation and Monte Carlo simulation with single-molecule quantum
  trajectories}}.
\newblock \emph{\bibinfo{journal}{Physical Review E}}
  \textbf{\bibinfo{volume}{90}}, \bibinfo{pages}{042719}
  (\bibinfo{year}{2014}).
\newblock
  \urlprefix\url{https://journals.aps.org/pre/abstract/10.1103/PhysRevE.90.042719}.

\bibitem{Cai2012}
\bibinfo{author}{Cai, J.}, \bibinfo{author}{Caruso, F.} \&
  \bibinfo{author}{Plenio, M.~B.}
\newblock \bibinfo{title}{{Quantum limits for the magnetic sensitivity of a
  chemical compass}}.
\newblock \emph{\bibinfo{journal}{Physical Review A}}
  \textbf{\bibinfo{volume}{85}}, \bibinfo{pages}{040304}
  (\bibinfo{year}{2012}).
\newblock
  \urlprefix\url{https://journals.aps.org/pra/abstract/10.1103/PhysRevA.85.040304}.

\bibitem{Fay2019}
\bibinfo{author}{Fay, T.~P.}, \bibinfo{author}{Lindoy, L.~P.},
  \bibinfo{author}{Manolopoulos, D.~E.} \& \bibinfo{author}{Hore, P.~J.}
\newblock \bibinfo{title}{{How quantum is radical pair magnetoreception?}}
\newblock \emph{\bibinfo{journal}{Faraday Discussions}}
  \textbf{\bibinfo{volume}{221}}, \bibinfo{pages}{77--91}
  (\bibinfo{year}{2019}).
\newblock \urlprefix\url{https://doi.org/10.1039/C9FD00049F}.

\bibitem{Carrillo2015}
\bibinfo{author}{Carrillo, A.}, \bibinfo{author}{Cornelio, M.~F.} \&
  \bibinfo{author}{de~Oliveira, M.~C.}
\newblock \bibinfo{title}{{Environment-induced anisotropy and sensitivity of
  the radical pair mechanism in the avian compass}}.
\newblock \emph{\bibinfo{journal}{Physical Review E}}
  \textbf{\bibinfo{volume}{92}}, \bibinfo{pages}{012720}
  (\bibinfo{year}{2015}).
\newblock
  \urlprefix\url{https://journals.aps.org/pre/abstract/10.1103/PhysRevE.92.012720}.

\bibitem{Wong2021}
\bibinfo{author}{Wong, S.~Y.}, \bibinfo{author}{Solov'yov, I.~A.},
  \bibinfo{author}{Hore, P.~J.} \& \bibinfo{author}{Kattnig, D.~R.}
\newblock \bibinfo{title}{{Nuclear polarization effects in cryptochrome-based
  magnetoreception}}.
\newblock \emph{\bibinfo{journal}{The Journal of Chemical Physics}}
  \textbf{\bibinfo{volume}{154}}, \bibinfo{pages}{035102}
  (\bibinfo{year}{2021}).
\newblock \urlprefix\url{https://aip.scitation.org/doi/abs/10.1063/5.0038947}.

\bibitem{Lewis2014}
\bibinfo{author}{Lewis, A.~M.}, \bibinfo{author}{Manolopoulos, D.~E.} \&
  \bibinfo{author}{Hore, P.~J.}
\newblock \bibinfo{title}{{Asymmetric recombination and electron spin
  relaxation in the semiclassical theory of radical pair reactions}}.
\newblock \emph{\bibinfo{journal}{The Journal of Chemical Physics}}
  \textbf{\bibinfo{volume}{141}}, \bibinfo{pages}{044111}
  (\bibinfo{year}{2014}).
\newblock \urlprefix\url{https://aip.scitation.org/doi/abs/10.1063/1.4890659}.

\bibitem{Keens2018}
\bibinfo{author}{Keens, R.~H.}, \bibinfo{author}{Bedkihal, S.} \&
  \bibinfo{author}{Kattnig, D.~R.}
\newblock \bibinfo{title}{{Magnetosensitivity in dipolarly coupled three-Spin
  systems}}.
\newblock \emph{\bibinfo{journal}{Physical Review Letters}}
  \textbf{\bibinfo{volume}{121}}, \bibinfo{pages}{096001}
  (\bibinfo{year}{2018}).
\newblock
  \urlprefix\url{https://journals.aps.org/prl/abstract/10.1103/PhysRevLett.121.096001}.

\bibitem{Kattnig2017}
\bibinfo{author}{Kattnig, D.~R.} \& \bibinfo{author}{Hore, P.~J.}
\newblock \bibinfo{title}{{The sensitivity of a radical pair compass
  magnetoreceptor can be significantly amplified by radical scavengers}}.
\newblock \emph{\bibinfo{journal}{Scientific Reports}}
  \textbf{\bibinfo{volume}{7}}, \bibinfo{pages}{1--12} (\bibinfo{year}{2017}).
\newblock \urlprefix\url{https://www.nature.com/articles/s41598-017-09914-7}.

\bibitem{Dellis2012}
\bibinfo{author}{Dellis, A.~T.} \& \bibinfo{author}{Kominis, I.~K.}
\newblock \bibinfo{title}{{The quantum Zeno effect immunizes the avian compass
  against the deleterious effects of exchange and dipolar interactions}}.
\newblock \emph{\bibinfo{journal}{Biosystems}} \textbf{\bibinfo{volume}{107}},
  \bibinfo{pages}{153--157} (\bibinfo{year}{2012}).
\newblock
  \urlprefix\url{https://www.sciencedirect.com/science/article/pii/S0303264711001894?via%3Dihub}.

\bibitem{Pedersen2016}
\bibinfo{author}{Pedersen, J.~B.}, \bibinfo{author}{Nielsen, C.} \&
  \bibinfo{author}{Solov'yov, I.~A.}
\newblock \bibinfo{title}{{Multiscale description of avian migration: from
  chemical compass to behaviour modeling}}.
\newblock \emph{\bibinfo{journal}{Scientific Reports}}
  \textbf{\bibinfo{volume}{6}}, \bibinfo{pages}{1--12} (\bibinfo{year}{2016}).
\newblock \urlprefix\url{https://www.nature.com/articles/srep36709}.

\bibitem{Babcock2020}
\bibinfo{author}{Babcock, N.~S.} \& \bibinfo{author}{Kattnig, D.~R.}
\newblock \bibinfo{title}{{Electron–electron dipolar interaction poses a
  challenge to the radical pair mechanism of magnetoreception}}.
\newblock \emph{\bibinfo{journal}{The Journal of Physical Chemistry Letters}}
  \textbf{\bibinfo{volume}{11}}, \bibinfo{pages}{2414--2421}
  (\bibinfo{year}{2020}).
\newblock
  \urlprefix\url{https://pubs.acs.org/doi/full/10.1021/acs.jpclett.0c00370}.

\bibitem{Babcock2021}
\bibinfo{author}{Babcock, N.~S.} \& \bibinfo{author}{Kattnig, D.~R.}
\newblock \bibinfo{title}{{Radical scavenging could answer the challenge posed
  by electron–electron dipolar interactions in the cryptochrome compass
  model}}.
\newblock \emph{\bibinfo{journal}{JACS Au}} \textbf{\bibinfo{volume}{14}},
  \bibinfo{pages}{jacsau.1c00332} (\bibinfo{year}{2021}).
\newblock \urlprefix\url{https://pubs.acs.org/doi/full/10.1021/jacsau.1c00332}.

\bibitem{Adams2018}
\bibinfo{author}{Adams, B.}, \bibinfo{author}{Sinayskiy, I.} \&
  \bibinfo{author}{Petruccione, F.}
\newblock \bibinfo{title}{{An open quantum system approach to the radical pair
  mechanism}}.
\newblock \emph{\bibinfo{journal}{Scientific Reports}}
  \textbf{\bibinfo{volume}{8}}, \bibinfo{pages}{1--10} (\bibinfo{year}{2018}).
\newblock \urlprefix\url{https://doi.org/10.1038/s41598-018-34007-4}.

\bibitem{Berkelbach2020}
\bibinfo{author}{Berkelbach, T.~C.} \& \bibinfo{author}{Thoss, M.}
\newblock \bibinfo{title}{{Special topic on dynamics of open quantum systems}}.
\newblock \emph{\bibinfo{journal}{The Journal of Chemical Physics}}
  \textbf{\bibinfo{volume}{152}}, \bibinfo{pages}{020401}
  (\bibinfo{year}{2020}).
\newblock \urlprefix\url{http://aip.scitation.org/doi/10.1063/1.5142731}.

\bibitem{Tanimura2020}
\bibinfo{author}{Tanimura, Y.}
\newblock \bibinfo{title}{{Numerically ``exact" approach to open quantum
  dynamics: The hierarchical equations of motion (HEOM)}}.
\newblock \emph{\bibinfo{journal}{The Journal of Chemical Physics}}
  \textbf{\bibinfo{volume}{153}}, \bibinfo{pages}{020901}
  (\bibinfo{year}{2020}).
\newblock \urlprefix\url{https://doi.org/10.1063/5.0011599}.

\bibitem{Suess2014}
\bibinfo{author}{Suess, D.}, \bibinfo{author}{Eisfeld, A.} \&
  \bibinfo{author}{Strunz, W.~T.}
\newblock \bibinfo{title}{{Hierarchy of stochastic pure states for open quantum
  system dynamics}}.
\newblock \emph{\bibinfo{journal}{Physical Review Letters}}
  \textbf{\bibinfo{volume}{113}}, \bibinfo{pages}{150403}
  (\bibinfo{year}{2014}).
\newblock
  \urlprefix\url{https://journals.aps.org/prl/abstract/10.1103/PhysRevLett.113.150403}.

\bibitem{Hartmann2017}
\bibinfo{author}{Hartmann, R.} \& \bibinfo{author}{Strunz, W.~T.}
\newblock \bibinfo{title}{{Exact open quantum system dynamics using the
  hierarchy of pure states (HOPS)}}.
\newblock \emph{\bibinfo{journal}{Journal of Chemical Theory and Computation}}
  \textbf{\bibinfo{volume}{13}}, \bibinfo{pages}{5834--5845}
  (\bibinfo{year}{2017}).
\newblock \urlprefix\url{https://pubs.acs.org/doi/10.1021/acs.jctc.7b00751}.

\bibitem{Breuer2015}
\bibinfo{author}{Breuer, H.-P.}, \bibinfo{author}{Laine, E.-M.},
  \bibinfo{author}{Piilo, J.} \& \bibinfo{author}{Vacchini, B.}
\newblock \bibinfo{title}{{Colloquium: Non-Markovian dynamics in open quantum
  systems}}.
\newblock \emph{\bibinfo{journal}{Reviews of Modern Physics}}
  \textbf{\bibinfo{volume}{88}}, \bibinfo{pages}{021002}
  (\bibinfo{year}{2016}).
\newblock
  \urlprefix\url{https://link.aps.org/doi/10.1103/RevModPhys.88.021002}.

\bibitem{DeVega2017}
\bibinfo{author}{{De Vega}, I.} \& \bibinfo{author}{Alonso, D.}
\newblock \bibinfo{title}{{Dynamics of non-Markovian open quantum systems}}.
\newblock \emph{\bibinfo{journal}{Reviews of Modern Physics}}
  \textbf{\bibinfo{volume}{89}}, \bibinfo{pages}{015001}
  (\bibinfo{year}{2017}).
\newblock
  \urlprefix\url{https://journals.aps.org/rmp/abstract/10.1103/RevModPhys.89.015001}.

\bibitem{Wu2020}
\bibinfo{author}{Wu, K.-D.} \emph{et~al.}
\newblock \bibinfo{title}{{Detecting non-Markovianity via quantified coherence:
  theory and experiments}}.
\newblock \emph{\bibinfo{journal}{npj Quantum Information}}
  \textbf{\bibinfo{volume}{6}}, \bibinfo{pages}{1--7} (\bibinfo{year}{2020}).
\newblock \urlprefix\url{https://www.nature.com/articles/s41534-020-0283-3}.

\end{thebibliography}

\end{document}